\documentclass[12pt]{article} 
\usepackage{epsfig}
\begin{document}

\begin{center}
CONTRIBUTION OF THE "HANGED" DIAGRAMS \\
 INTO THE REACTION $np \rightarrow np \pi^+ \pi^-$ 
\\
\vspace{0.3cm}
A.P. Jerusalimov
\\
JINR, Dubna, Moscow region, 141980, Russia 
\end{center}

\vspace{0.3cm}
\begin{abstract}
  The contribution of "hanged" diagrams into the reaction $np \rightarrow np \pi^+ \pi^-$
was considered. It was shown that taking into account of these diagrams permits to get
better description of the effective mass spectrum of $\pi^+\pi^-$-combinations.
\end{abstract}

\vspace{1.0cm}

  In paper\cite{OPERnp2pi} it was studied the mechanism of the reaction $np \rightarrow 
np \pi^+ \pi^-$ at intermediate energies ($1.73 < P_0<5.2 GeV/c$). It was shown that the 
main contribution made the diagrams with exchange by reggeized pi-meson (Fig.5 and Fig.7
in~\cite{OPERnp2pi}). Moreover it is necessary to take into account also the diagrams of
one-baryon exchange (Fig.10 in~\cite{OPERnp2pi}) at energies below 3 GeV.
  A good description of the characteristics of the reaction $np \rightarrow np \pi^+ \pi^-$
was obtained at he considered energy region. So-called "hanged" diagrams of pi-meson 
exchange (similar to shown in Fig.1) was not considered because their contribution did 
not exceed $2\%$ at $P_0=5.20 GeV/c$.

\begin{figure}[h]
\hspace{3.5cm}
\includegraphics[width=0.5\textwidth]{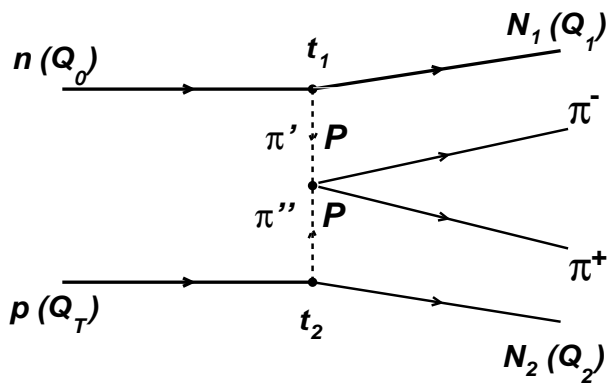}
\caption{So-called "hanged" diagrams for the reaction $np \rightarrow np \pi^+ \pi^-$
including $\pi$-meson and pomeron (P) exchanges.}
\label{Fig1}
\end{figure}

 Some exceeding of experimental distribution above the theoretical curve in the effective 
mass spectrum of $\pi^+\pi^-$ - combinations at $P_0=1.73$ GeV/c (Fig.11b 
in~\cite{OPERnp2pi}) was considered as a fluctuation at small statistics.

 But the data of the reaction $np \rightarrow np \pi^+ \pi^-$ at $T_{kin}=1.25
GeV$~\cite{np2piH}
(~$10^7$ events) obtained at HADES set-up~~\cite{HADES} shown that the bump in the 
effective mass spectrum of $\pi^+\pi^-$ - combinations is statistically significant and 
therefore has a dynamic nature. Then it was decided to study the contribution of the 
"hanged" diagrams into the reaction $np \rightarrow np \pi^+ \pi^-$ more detailed.

  Let us concider the structure of these "hanged" diagrams.

  The matrix element of the "hanged" diagrams being the result of $\pi$-meson exchange is 
written in the following form:
\begin{equation}
\label{Tpi}
  T_{\pi} = G \,\bar u(Q_1)\gamma_5 u(Q_0) 
  \; \frac{F_{\pi}(S_{N_1 \pi \pi},t_1,t_2)}{t_1-m^2_{\pi}}
  \; T_{\pi\pi}(S_{\pi\pi},t_1,t_2)  
  \;\frac{F_{\pi}(S_{N_2 \pi \pi},t_1,t_2)}{t_2-m^2_{\pi}}
  \; G \, \bar u(Q_2)\gamma_5 u(Q_T)
\end{equation}

 where $\;\bar u(Q_i)\gamma_5 u(Q_j)$ - vertex functions,\\
\hspace*{1.8cm} $F_{\pi}$ - formfactors in the form taken from \cite{OPERnp2pi},\\
\hspace*{1.8cm} $T_{\pi\pi}$ - off shell amplitude of elastic $\pi\pi$-scattering 
                ~(\cite{pipiscat1},\cite{pipiscat2}),\\
\hspace*{1.8cm} $G$ the constant of strong interaction ($G^2/4pi=14.6$),\\
\hspace*{1.8cm} $t_1=(Q_0-Q_1)^2$,\\
\hspace*{1.8cm} $t_2=(Q_T-Q_2)^2$,\\
\hspace*{1.8cm} $S_{N\pi\pi}$: $(Q_1+q_1+q_2)^2$ and $(Q_2+ q_1+q_2)^2$,\\
\hspace*{1.8cm} $S_{pipi}=()q_1+q_2)^2$.\\

  The corresponding matrix element for the pomeron (P) exchange is written in the form:
\begin{equation}
\label{TP}
  T_P = g_P(t_1) \,
  F_P(S_{N_1 \pi \pi},t_1,t_2) \;
  \; T_{\pi\pi}^{0,0}(S_{\pi\pi},t_1,t_2) \;  
  \;F_P(S_{N_2 \pi \pi},t_1,t_2) \,
  \; g_P(t_2)
\end{equation}

 where $\;g(t)$ - vertex functions~\cite{Nikitin},\\
\hspace*{1.8cm} $F_P$ - formfactors with parameters taken from~\cite{Nikitin},\\
\hspace*{1.8cm} $T_{\pi\pi}^{00}$ the S-wave (I=0, L=0) amplitude of elastic $\pi\pi$
  -scattering~(\cite{pipiscat1},\cite{pipiscat2}),\\

  Squared matrix element of the reaction $np \rightarrow np \pi^+ \pi^-$ 
was written in in the form:
$$T_h=|T_{\pi^0}|^2+|T_{\pi^{\pm}}|^2+|T_P|^2$$
neglecting the interference of the diagrams for the present.

  The results of the calculations for the reactions $np \rightarrow np \pi^+ \pi^-$ 
at $P_0=1.73$ GeV/c are shown in Fig.2. One can see that taking into account the "hanged"
diagrams permits to get the noticeably better description of the $\pi^+\pi^-$ masses 
close to 300 $MeV/c^2$. 

  The same calculations were carried out also for the reaction 
$np \rightarrow np \pi^+ \pi^-$ at $T_{kin}=1.25$ GeV to compare with the data obtained
from HADES set-up. The result is presented in Fig.3.



\begin{figure}[h]
\hspace{2.0cm}
\includegraphics[width=0.6\textwidth]{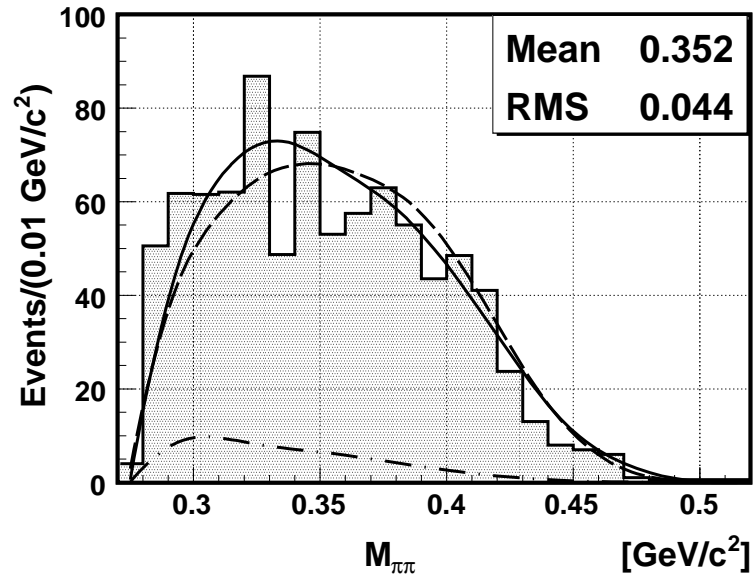}
\caption{The spectrum of effective masses of $\pi^+\pi^-$-combinations from the reaction 
  $np \rightarrow np \pi^+ \pi^-$ at $P_0=1.73$ GeV/c. Solid line - the result of taking
  into account "hanged" diagrams, dashed line - without "hanged" diagrams, dash-dotted
  line - the contribution of "hanged" diagrams.}
\label{Fig2}
\end{figure}

\begin{figure}[h]
\hspace{2.0cm}
\includegraphics[width=0.6\textwidth]{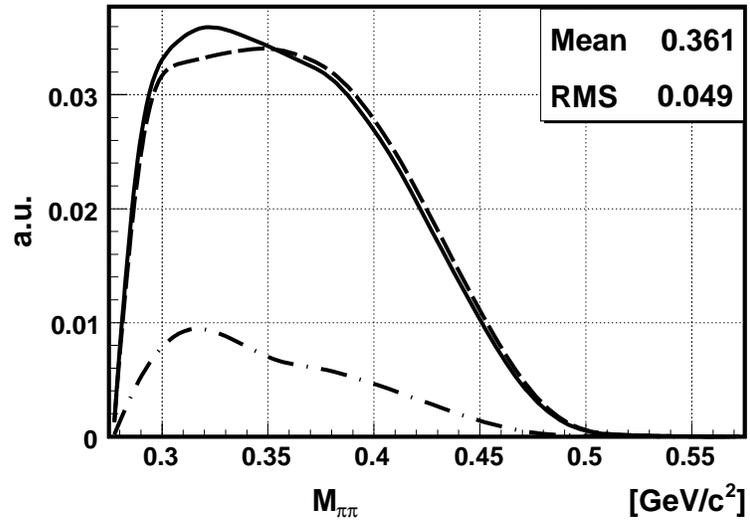}
\caption{The spectrum of effective masses of $\pi^+\pi^-$-combinations from the reaction 
  $np \rightarrow np \pi^+ \pi^-$ at $T_{kin}=1.25$ GeV. The notations are the same as in
  Fig.2.}
\label{Fig3}
\end{figure}

\end{document}